\documentclass[12pt]{iopart}
\usepackage{iopams}
\usepackage{cite}
\usepackage{multirow}
\usepackage{colortbl}

\usepackage{color}

\usepackage{appendix}
\usepackage[utf8]{inputenc}
\usepackage{graphicx}
\usepackage{graphics}
\usepackage[labelfont=bf]{caption}
\usepackage{subcaption}
\PassOptionsToPackage{hyphens}{url}
\usepackage[hyphens]{url}
\usepackage[colorlinks=true,urlcolor=blue,hidelinks,pdftex]{hyperref}
\usepackage{xurl}

\usepackage{tikz,xcolor}

\definecolor{lime}{HTML}{A6CE39}
\DeclareRobustCommand{\orcidicon}{
	\begin{tikzpicture}
	\draw[lime, fill=lime] (0,0) 
	circle [radius=0.16] 
	node[white] {{\fontfamily{qag}\selectfont \tiny ID}};
	\draw[white, fill=white] (-0.0625,0.095) 
	circle [radius=0.007];
	\end{tikzpicture}
	\hspace{-2mm}
}

\foreach \x in {A, ..., Z}{\expandafter\xdef\csname orcid\x\endcsname{\noexpand\href{https://orcid.org/\csname orcidauthor\x\endcsname}
			{\noexpand\orcidicon}}
}

\DeclareGraphicsExtensions{.png,.jpg,.JPEG,.bmp,.pdf,.eps,.tiff}

\usepackage{hyperref}

\begin{document}
\noindent\rule{\textwidth}{1pt}

\title[Two blocks connected by a string with variable tension: A dynamic case.]{Two blocks connected by a string with variable tension: A dynamic case.}
%\author{H. J. Herrera Su\'{a}rez}
%    \affiliation{Facultad de Ciencias Naturales y Matem\'{a}ticas, Universidad de Ibagu\'{e}, Ibagu\'{e}-Colombia, Carrera 22 Calle 67, barrio Ambal\'{a} \\ ORCID: 0000-0003-1273-7037}
%    \email{hernan.herrera@unibague.edu.co}
%\author{ M. Machado-Higuera}https://www.leybold-shop.com/physics/physics-equipment/systems/cassy/sensor-boxes-sensors/physics/timer-s-524074.html
%    \affiliation{Facultad de Ciencias Naturales y Matem\'{a}ticas, Universidad de Ibagu\'{e}, Ibagu\'{e}-Colombia, Carrera 22 Calle 67, barrio Ambal\'{a}.}
%    \email{maximiliano.machado@unibague.edu.co}
%\author{J. H. Mu\~{n}oz}%
%        \affiliation{Departamento de F\'{\i}sica, Universidad del Tolima, Ibagu\'{e}-Colombia.}
%    \email{jhmunoz@ut.edu.co}

\author{H. J. Herrera-Su\'{a}rez $^{1,a}$ \orcidA{}  \& M. Machado-Higuera $^{1,b}$ \orcidB{} \& J. H. Mu\~{n}oz $^{2,c}$ \orcidC{}} 
\address{$^1$ Universidad de Ibagu\'{e}, Facultad de Ciencias Naturales y 
Matem\'{a}ticas, Ibagu\'{e}-Colombia, Carrera 22 Calle 67, barrio Ambal\'{a}}

\address{$^2$ Universidad del Tolima, Departamento de F\'{\i}sica,  
Ibagu\'{e}-Colombia, Barrio Santa Helena Parte Alta.}

\ead{$^a$ hernan.herrera@unibague.edu.co \& $^b$ maximiliano.machado@unibague.edu.co \& 
$^c$ jhmunoz@ut.edu.co}
\vspace{10pt}

\begin{indented}
\item[] April 2020
\end{indented}

\begin{abstract}
In this paper, the dynamic case of a system made up of two blocks connected by a string over a smooth pulley is revisited. One mass lies on a horizontal surface without friction, and the other mass has a vertical displacement. The motion equation is obtained and its solution is determined using the \textit{Mathematica} package. Also  an experimental montage for this system is made and   experimental data for the vertical position $y$ in function of the time $t$ are obtained using a Data Acquisition System   and the \textit{Tracker} video analysis. The relation $y$ vs $t$ can be represented by a polynomial of degree six.  An average relative error of $3.61 (10.14) \%$ is obtained  between the theoretical results acquired with  \textit{Mathematica} and the data taken from the \textit{Tracker} (Data Acquisition System). 
\end{abstract}

%\maketitle

\section{Introduction}

In this work, a system of two blocks of masses $m_{1}$ and $m_{2}$ connected by a string over a smooth pulley and subjected to variable acceleration is studied. The mass $m_{2}$ is suspended, and the mass $m_{1}$ moves on a horizontal surface without a coefficient of  dynamic friction. The string is extensionless, and  uniform, and its  mass is negligible. Figure 1  shows the forces acting on this system.

{Unlike other systems made up of two masses tied to a rope that passes through a pulley (such as those shown in Figure 2), this problem is significant because the tension $T$, the normal force $N$ and the acceleration are not constant because they depend on the variation of the angle $\theta$. This system is a more complex progression of the four common configurations shown in Figure 2}.

The static situation of this problem has been studied previously in some fundamental physics textbooks \cite{Serway2014, Serway2014a, Gonz, Ohanian, Hibbeler, Riley}, in several papers \cite{Lerman, Mak, Sutt, van, Leonard, DHH} and on  the website of A. Franco \cite{Franco}. The dynamic case, in which there is no  friction between the horizontal surface and the mass $m_1$, was proposed as a problem in  Serway-Jewett's physics textbook   \cite{Serway2014} and its solution appears in the Instructor's Solution Manual of the same author \cite{Serway2014a}. In this paper,  this case is revisited with the purpose of performing a {complete} analysis to this system. The motion equation is explicitly obtained,  and its solution is found  using the \textit{Mathematica} package. An experimental montage is made,  and  experimental data for the position in function of the time for the mass $m_{2}$ are got using a Data Acquisition System (DAS)  and the \textit{Tracker} video analysis. The experimental results are then compared  with the theoretical solution.

\begin{figure}
    \centering
    \includegraphics[scale=0.05]{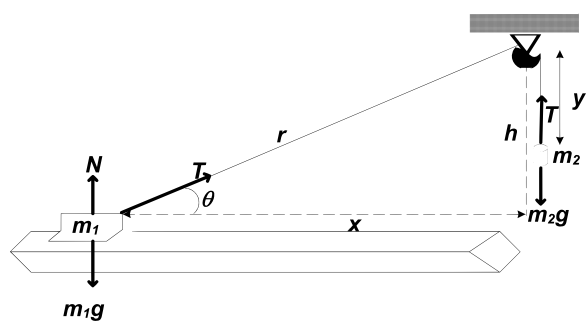}
    \caption{Two blocks tied to an extensionless string.}
    \label{fig:1}
\end{figure}

\begin{figure}
    \begin{subfigure}[b]{.4\linewidth}
        \centering
        \includegraphics[width=\linewidth]{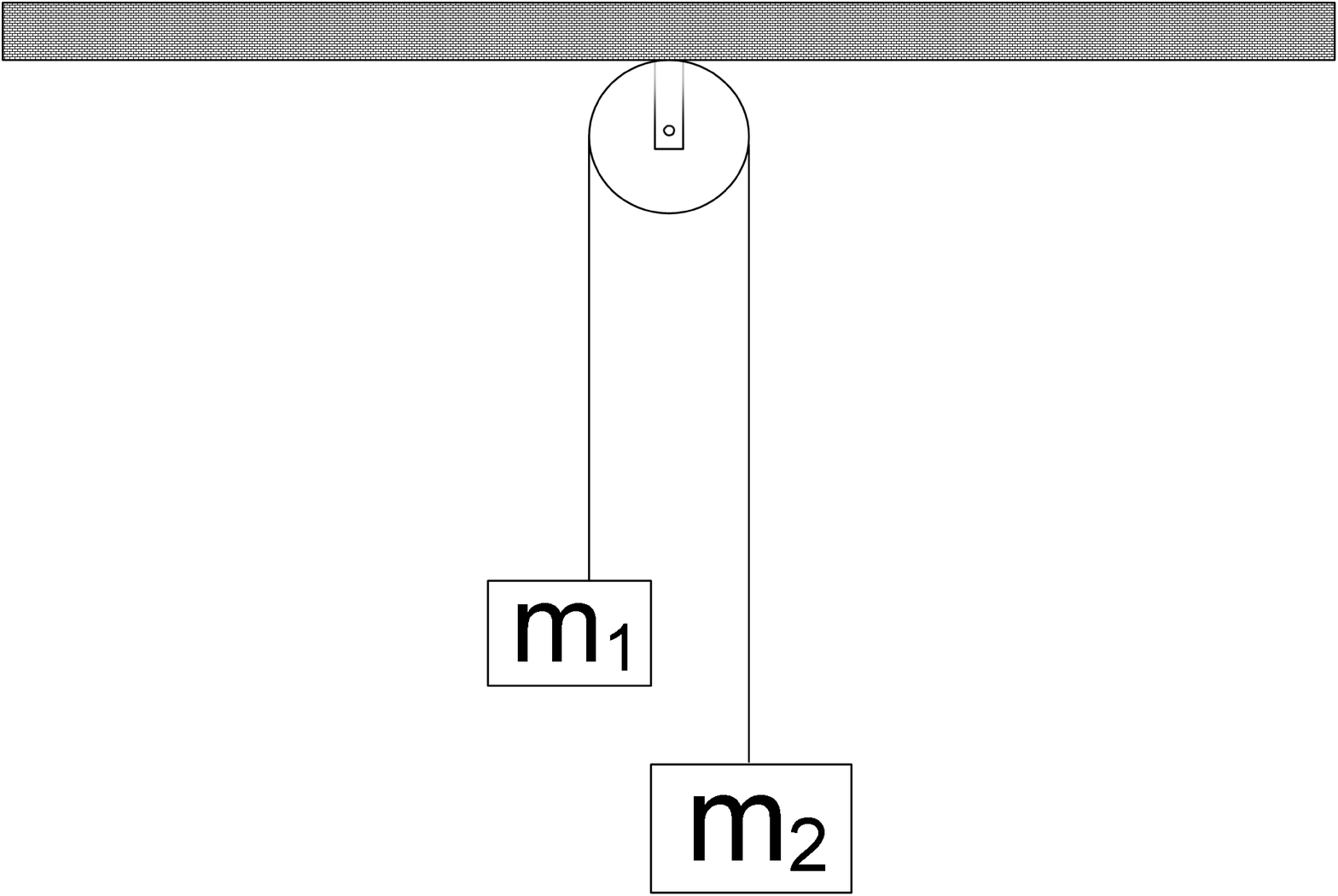}
        
    \end{subfigure}
    \begin{subfigure}[b]{.4\linewidth}
        \centering
        \includegraphics[width=\linewidth]{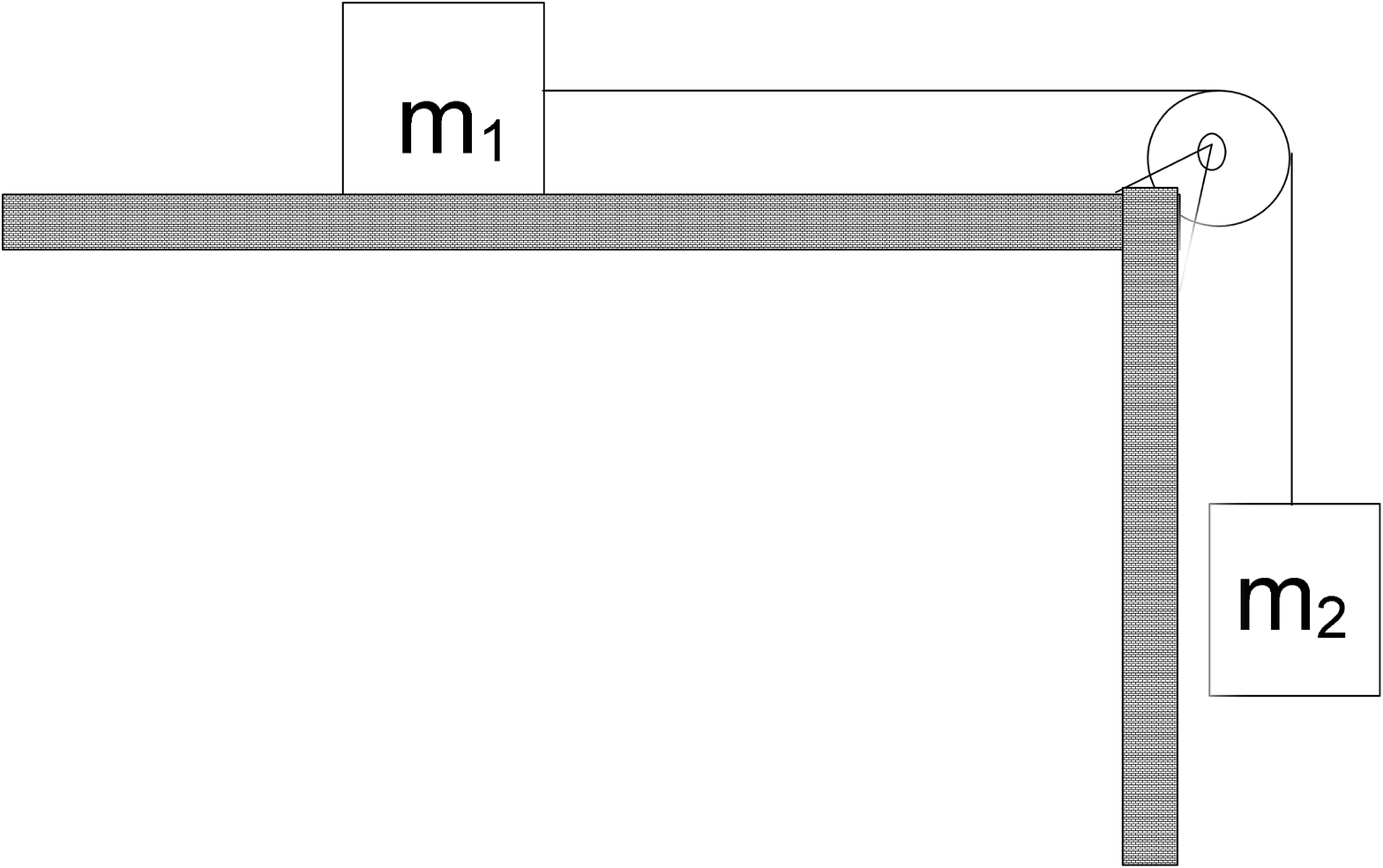}
       
    \end{subfigure}
    
    \begin{subfigure}[b]{.4\linewidth}
        \centering
        \includegraphics[width=\linewidth]{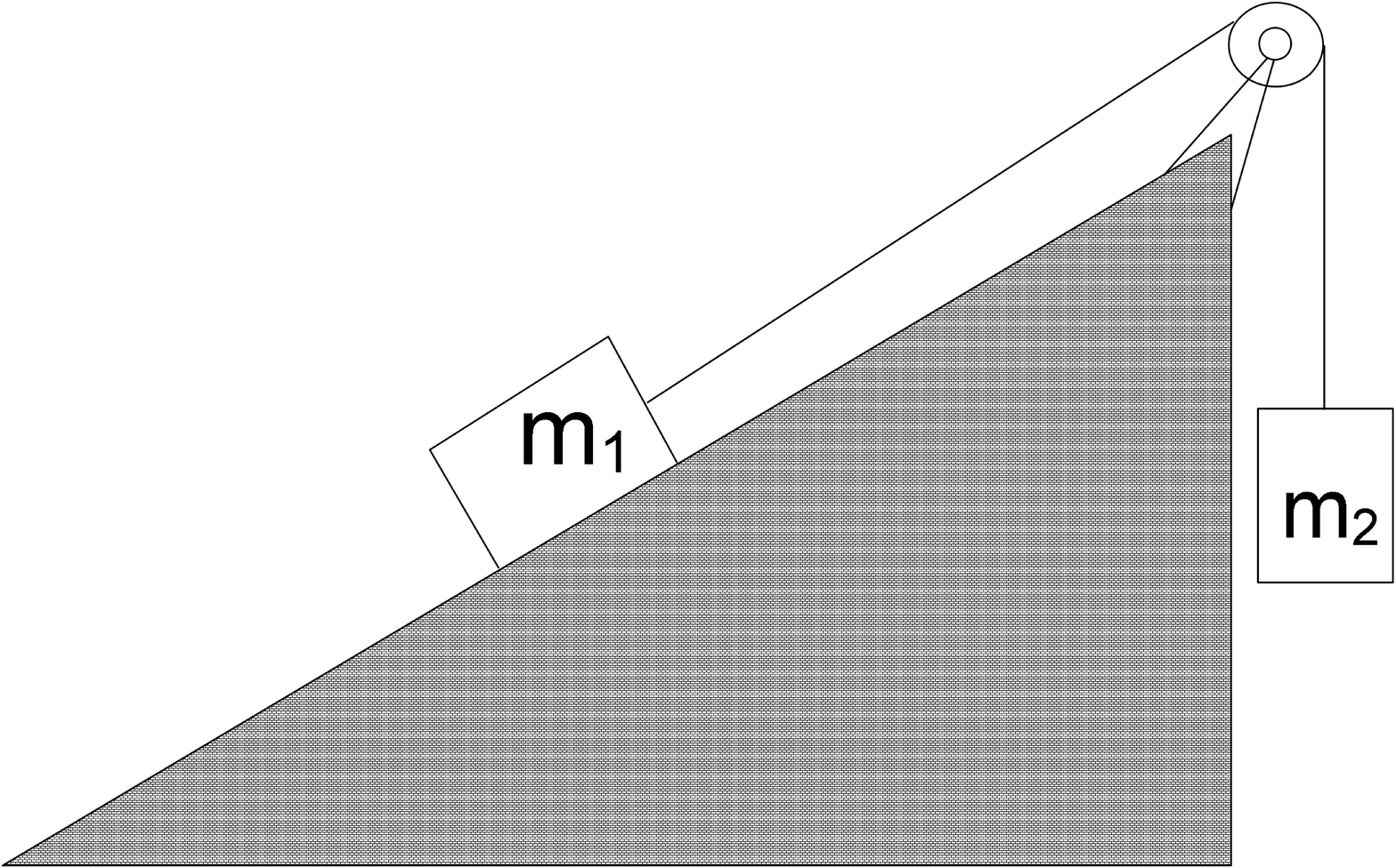}
        \
    \end{subfigure}
    \begin{subfigure}[b]{.4\linewidth}
        \centering
        \includegraphics[width=\linewidth]{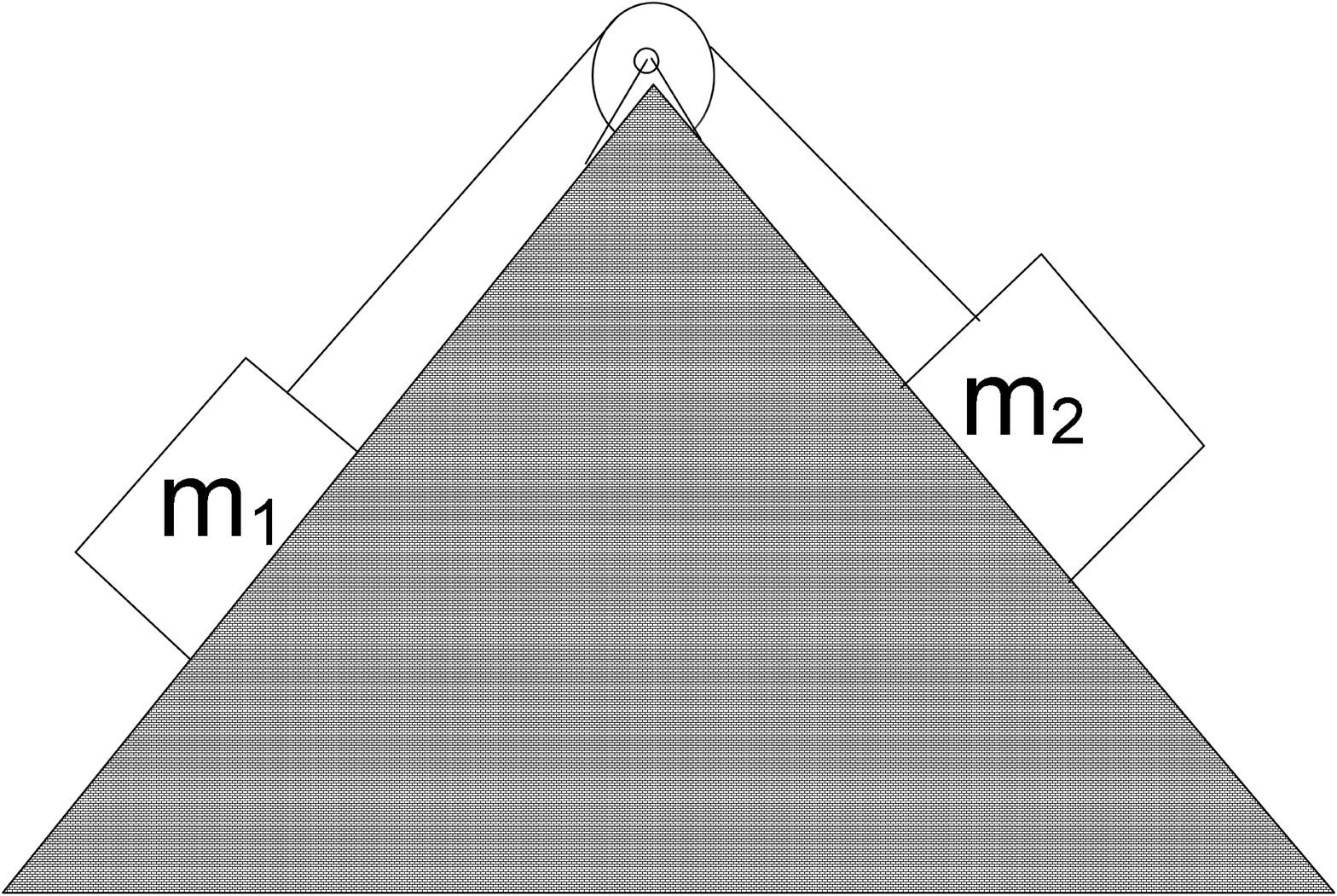}
        
    \end{subfigure}
    \caption{{Systems made up of two masses tied to a rope that passes through a pulley. The tension and the acceleration are constant.}  }\label{fig:2a_2b}
\end{figure}

This paper is organized as follows. In Section 2, the experimental arrangement built by the authors to represent the system shown in \Fref{fig:1} is displayed; in Section 3, the theoretical analysis is presented, and the results are  briefly discussed in Section 4. Finally, in Section 5, some concluding remarks are given.

\section{Experimental arrangement}

\begin{figure}[htb]
    \centering
    \includegraphics[scale=0.4]{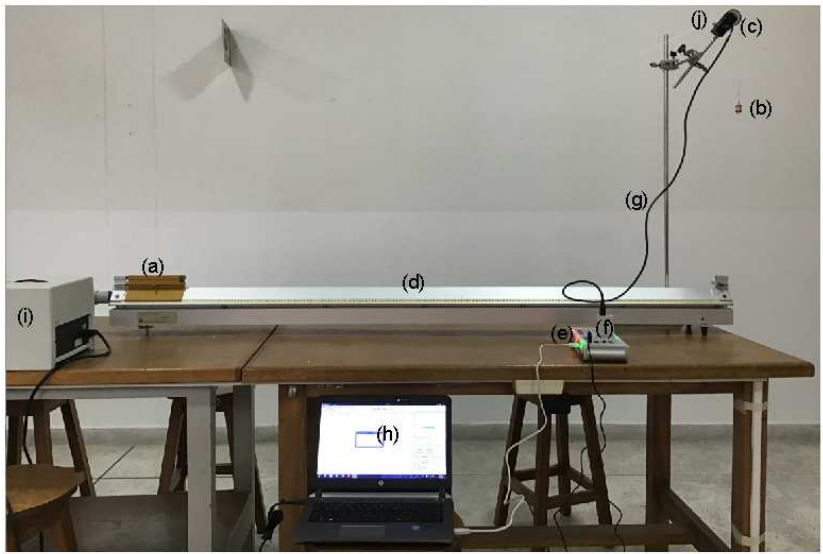}
    \caption{Experimental arrangement: (a) mass $m_{1}$; (b) mass $m_{2}$; (c) combination spoked wheel; (d) air track; (e) sensor CASSY 2; (f) timer S; (g) multi core cable, 6 pole, 1.5 m; (h) computer; (i) air supply, and (j) combination light barrier.}
    \label{fig:2}
\end{figure}

\Fref{fig:2} shows the experimental arrangement used to represent the system in Figure $1$. The two masses are 
{$m_{1}=574.12$ g and $m_{2}= 100$ g (measured with a WTC 2000 precision balance \cite{WTC})}, and the length of the string is $l=r+y=2.09$ m and $h=0.98$ m. The mass  $m_{2}$ moves down in the vertical direction, and the mass $m_{1}$ moves horizontally over  the linear air track, reference $11202-88$, of the company Phywe \cite{Phywe}. The DAS used is made up of: a sensor-CASSY 2, reference $524 013$ \cite{Phywea}; a timer S \cite{Phyweb};  a multi core cable, 6 pole, $1.5$ m, reference $501 16$ \cite{Phywec};  a combination spoked wheel, reference $337464$ \cite{Phywed}; and a combination light barrier, reference $337 462$ \cite{Phywee}, of the company Phywe, and a computer, which are fundamental to the acquisition of data. 

The air supply (see (i) in \Fref{fig:2}) in used to reduce the friction between the mass $m_{1}$ and the air track. The data for the variable $y$ in function of the time were registered  for the DAS with the initial condition $y(0)=-0.255$ m. The movement of the two masses was also recorded with  a smartphone, and the video was analized using the \textit{Tracker} video analysis \cite{Tracker}.

\section{Theoretical analysis}

The system displayed in \Fref{fig:1} is analyzed using  Newton's second law. The sum of the forces for the particle $m_{1}$ in the $x$ direction is

\begin{eqnarray}
T\cos \theta &=&m_{1}a_{x}; \label{eq:1}
\end{eqnarray}

and the sum of the forces for the mass $m_{2}$ in the $y$ direction gives

\begin{eqnarray}
T-m_{2}g &=&-m_{2}a_{y}. \label{eq:2}
\end{eqnarray}

From this equation, the tension $T$ is obtained and by substituting it  in  Equation (\ref{eq:1}) the following is got

\begin{eqnarray}
\left( m_{2}g-m_{2}a_{y}\right) \frac{x}{\sqrt{x^{2}+h^{2}}}=m_{1}a_{x},  \label{eq:3}
\end{eqnarray}%

where  $\cos \theta =x/r=x/\sqrt{x^{2}+h^{2}}$ (according to \Fref{fig:1}).

Considering  \Fref{fig:1}, the length $l$ of the rope is given by

\begin{equation}
l =\sqrt{x^{2}+h^{2}}+y. \label{eq:4}
\end{equation}

From this equation the following is obtained
\begin{equation}
x =\sqrt{(l-y)^{2}-h^{2}}. \label{eq:5}
\end{equation}
On the other hand, if {differentiated} with respect to time in Equation (\ref{eq:4}), the equation below is got
\begin{equation}
v_{y} =-xv_{x}({x^{2}+h^{2}})^{-\frac{1}{2}}, \label{eq:6}
\end{equation}
where $v_{x}=\frac{dx}{dt}$ and $v_{y}=\frac{dy}{dt}$ are the speeds of mass $m_{1}$ and $m_{2}$, respectively.

Now, by {differentiating} this expression with respect to time,  after replacing $x$ and $v_{x}$, by means of the Equations (\ref{eq:5}) and (\ref{eq:6}), respectively, the below expression is got
\begin{eqnarray}
a_{x} &=&-\left( \frac{h^{2}v_{y}^{2}}{\left( \left( l-y\right)
^{2}-h^{2}\right) ^{\frac{3}{2}}}+\frac{\left( l-y\right) a_{y}}{\left(
\left( l-y\right) ^{2}-h^{2}\right) ^{\frac{1}{2}}}\right),  \label{eq:7}
\end{eqnarray}

where $a_{x}=\frac{dv_{x}}{dt}$ and $a_{y}=\frac{dv_{y}}{dt}$ are the accelerations of mass $m_{1}$ and $m_{2}$, respectively.
If   $v_{y}=0$, then $a_{x}=-ua_{y}$ with $u=(l-y)({(l-y)^2{-h^2}})^{-{\frac{1}{2}}}$.
This result agrees with the one presented in the references \cite{Serway2014,Serway2014a}.

By Replacing the last expression for $a_{x}$ in Equation (\ref{eq:3}), the following expression is  obtained:

\begin{eqnarray}
a_{y}=\frac{m_{2}g\left( \left( l-y\right) ^{2}-h^{2}\right) ^{2}+\left(
l-y\right) m_{1}h^{2}v_{y}^{2}}{\left( \left( l-y\right) ^{2}-h^{2}\right)
\left( m_{2}\left( \left( l-y\right) ^{2}-h^{2}\right) -m_{1}\left(
l-y\right) ^{2}\right) }. \label{eq:8}
\end{eqnarray}

This differential equation gives the equation motion for $y$. 

\section{Results and discussion}

The evolution of the variable $y$ in function of the time $t$ has been obtained via three different ways:  (i) employing the \textit{Tracker}, (ii) manipulating  the DAS, and (iii) using the \textit{Mathematica} package.

\begin{figure}[htb]
    \centering
    \includegraphics[scale=0.32]{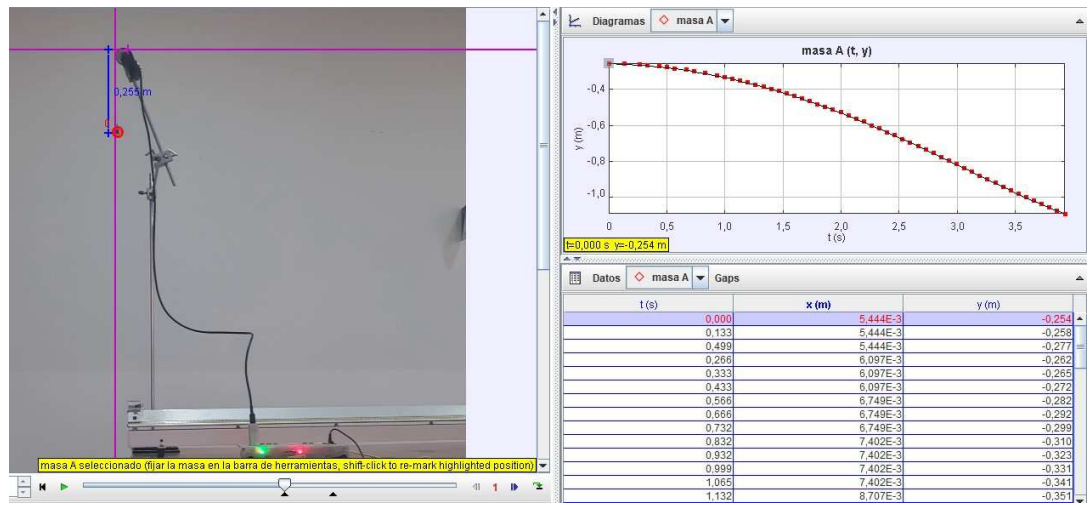}
    \caption{An image of the movement of the mass $m_{2}$ using the \textit{Tracker}.}
    \label{fig:3}
\end{figure}

\begin{figure}[htb]
    \centering
    \includegraphics[scale=0.4]{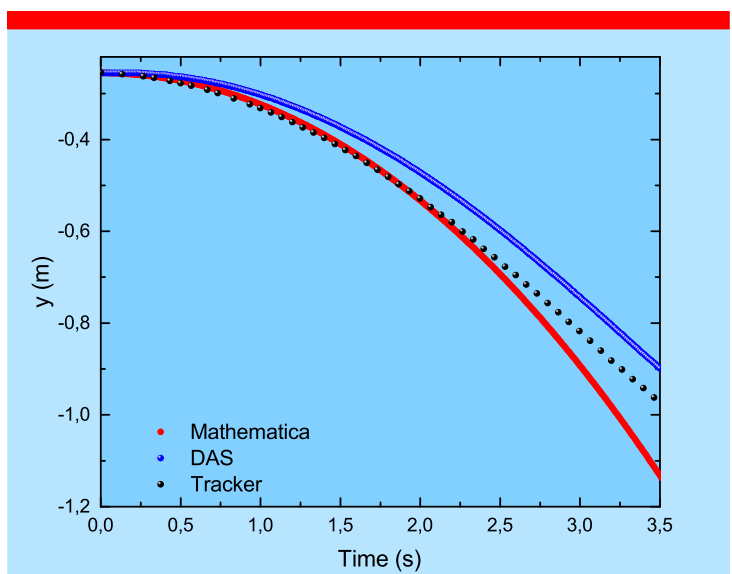}
    \caption{Graph of $y$ vs $t$ for the mass $m_2$. The solid black (blue)  circles represent the experimental data obtained with the \textit{Tracker} (DAS). The red circles represent the solution of the differential equation obtained with \textit{Mathematica}.}
    \label{fig:4}
\end{figure}

As mentioned in Section 2,  the movement of the masses $m_{1}$ and $m_{2}$ were recorded with a smartphone and the video was analyzed with the \textit{Tracker}. \Fref{fig:3} (see left side) shows a picture of the movement of the mass $m_{2}$ using this computational tool. The vertical displacement of the particle $m_{2}$ in function of the time,  $y$ vs $t$, is displayed in the top-right of this figure, and the data are shown in the bottom-right.  \Fref{fig:4} shows the graph of  $y$ vs $t$ obtained with the \textit{Tracker} (see the solid black circles). On the other hand, the software CASSY was also used to obtain $y$ vs $t$ from the DAS (see the  solid blue circles). 

The solution of the differential equation shown in Equation (\ref{eq:8}), was obtained by means of the \textit{Mathematica}, using the commands NDSolve and Plot[Evaluate[y[t] /. s, {t, 0, 1}]. This solution is displayed in \Fref{fig:4} (see the solid red  circles).

\Fref{fig:4} was obtained using the OriginPro package \cite{Origin}. This software allows the results obtained by the \textit{Tracker}, the DAS, and \textit{Mathematica} to be presented  in a single image. The experimental data confirm that the motion of the mass $m_{2}$ is neither  uniform nor uniformly accelerated.

\newpage

\begin{figure}[htb]
    \begin{subfigure}[b]{.48\linewidth}
        \centering
        \includegraphics[width=\linewidth]{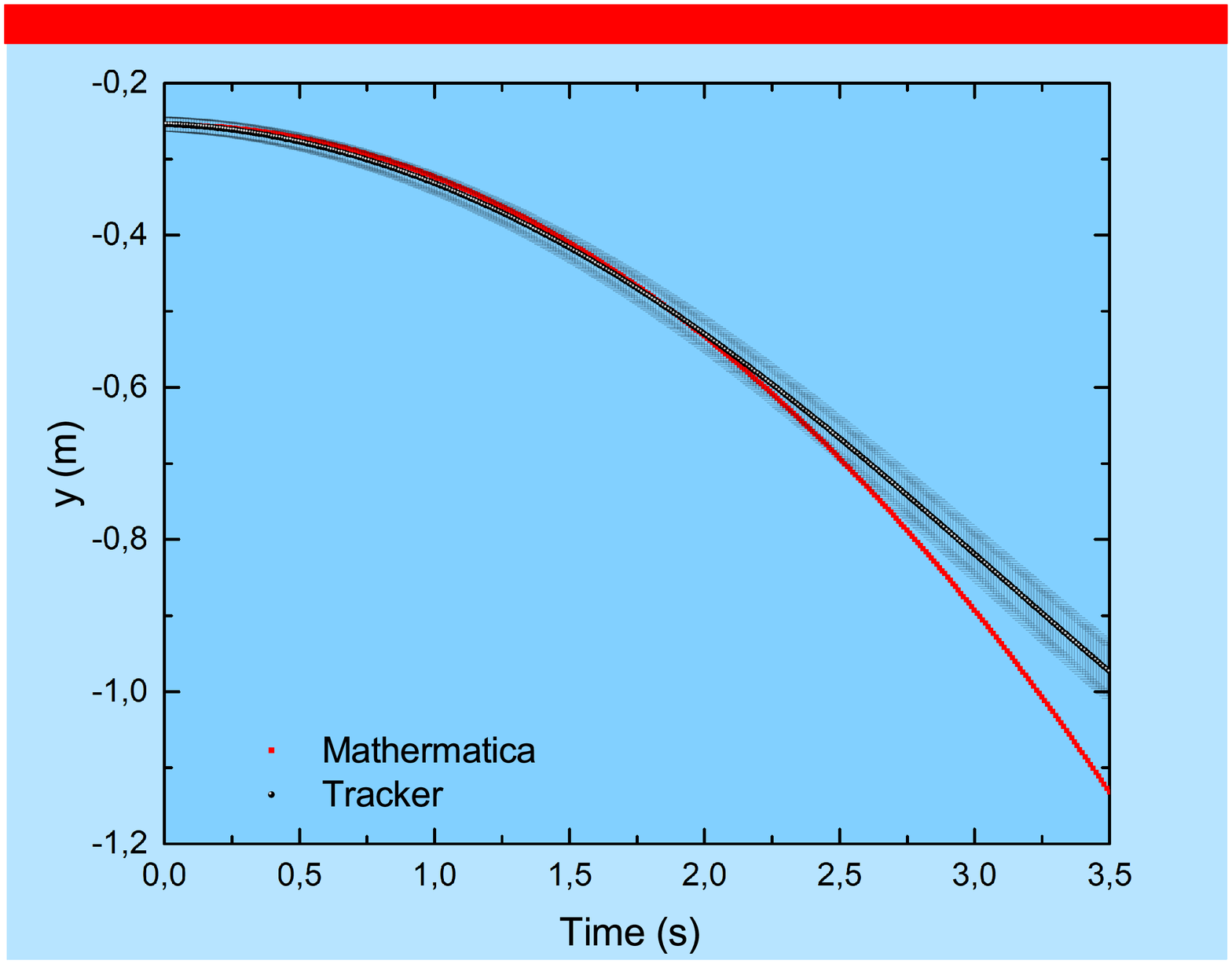}
        \caption{}\label{fig:5a}
    \end{subfigure}
    \begin{subfigure}[b]{.48\linewidth}
        \centering
        \includegraphics[width=\linewidth]{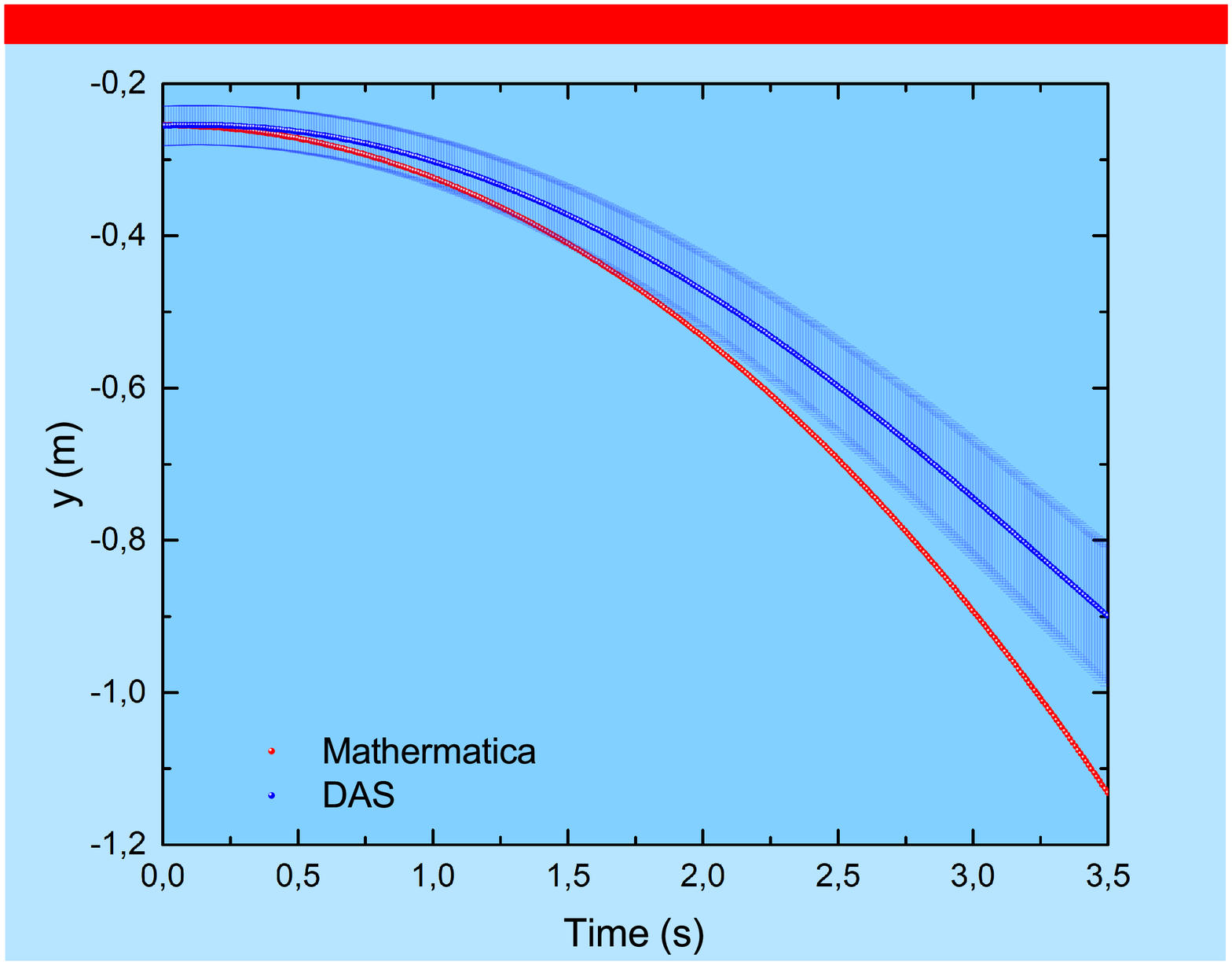}
        \caption{}\label{fig:5b}
    \end{subfigure}
    \caption{Comparison of the  obtained results  from  Mathematica  and  the Tracker (a) and the  DAS (b).}\label{fig:5a_5b}
\end{figure}

{The obtained data with the DAS and the Tracker and the acquired result with Mathematica, were fitted by means of the OriginPro. The best fit is given by a polynomial of degree six with a correlation coefficient of, approximately, 0.999. The corresponding expressions are display in the Table 1.  The curves and the errors shown in \Fref{fig:5a_5b} were obtained from these equations}. The results obtained  with \textit{Mathematica} and the  \textit{Tracker} (DAS) were compared (see \Fref{fig:5a} (\Fref{fig:5b})), giving  that  the average relative error of the theoretical results acquired with  \textit{Mathematica} in relation to the data taken with the \textit{Tracker} and the DAS is $3.61$ $\%$ and $10.14$ $\%$, respectively. 

{It is important to note that both measured curves indicate lower accelerations than the calculated indicating that there is some small friction. Further, the acceleration obtained with the DAS is slower than the one obtained with the Tracker probably because of cord slippage around the pulley. The curves show good agreement at the beginning and then diverge because the increasing angle is altering whatever friction there is.}

\newpage

\bigskip 

\begin{table}[htbp!]
  
\caption{\label{Table 1}  $y$ vs $t$ for the obtained data with the Tracker, the DAS and the theoretical solution given by Mathematica.}

\begin{indented}
    \item[] \begin{tabular}{ll}
   \rowcolor[rgb] {0.67, 0.9, 0.93} \br
    &  \\ 
\rowcolor[rgb] {0.67, 0.9, 0.93} \textit{Mathematica} & 
\begin{tabular}{ll}
\rowcolor[rgb] {0.67, 0.9, 0.93} $y=$ & $-1.45919\times 10^{-6}t^{6}+5.66164\times 10^{-5}t^{5}-$ \\ 
& $1.76691\times 10^{-4}t^{3}-0.06829t^{2}$ \\ 
& $+2.75608\times 10^{-5}t-0.255$%
\end{tabular}
\\ \hline
\rowcolor[rgb] {0.67, 0.9, 0.93} DAS & 
\begin{tabular}{ll}
\rowcolor[rgb] {0.67, 0.9, 0.93} $y=$ & $1.60902\times 10^{-4}t^{6}-0.00132t^{5}+0.00502t^{4}$ \\ 
& $-0.00811t^{3}-0.0576t^{2}+0.01529t-0.25567$%
\end{tabular}
\\ \hline
\rowcolor[rgb] {0.67, 0.9, 0.93} \textit{Tracker} & 
\begin{tabular}{ll}
\rowcolor[rgb] {0.67, 0.9, 0.93} $y=$ & $-1.75581\times 10^{-4}t^{6}+0.00213t^{5}-0.00773t^{4}$ \\ 
& $+0.01381t^{3}-0.07397t^{2}-0.0118t-0.25406$%
\end{tabular}
\\ \hline\hline
\end{tabular}
\end{indented}
\end{table}

\section{Concluding remarks}

The authors  studied the dynamic situation of a block of mass $m_{1}$ on a horizontal plane being pulled at an angle $\theta$ with the horizontal by a tension due to a suspended mass $m_{2}$, without considering the friction between $m_{1}$ and the horizontal plane. The experimental results were obtained using the \textit{Tracker} and the DAS shown in \Fref{fig:2}, and the theoretical solution   to the motion   equation was found using \textit{Mathematica}.

The movement of the mass $m_{2}$ along the y-axis is neither uniform  nor uniformly accelerated. The best fit for $y$ in function of the time is a polynomial of degree six. The theoretical prediction obtained with \textit{Mathematica} gives a better agreement with data taken using the \textit{Tracker} than using the DAS. 

This work is worthwhile because it allows the articulation between experiment and theory, facilitating students{'} understanding of the physics behind the theory. Besides, it allows one to check the experimental viability of theoretical problems proposed in physics textbooks.

%%%%%%%%%%%%%%%%%%%%%%%%%%%%%%%%%%%%%%%%%%

%%%%%%%%%%%%%%%%%%%%%%%%%%%%%%%%%%%%%%%%%
%\label{article@endpage}
%%%%%%%%%%%%%%%%%%%%%%%%%%%%%%%%%%%%%%%%%

\end{document}